\documentclass[npg,manuscript]{copernicus}
\nolinenumbers

\begin{document}

\title{An approach for projecting the timing of abrupt winter Arctic sea ice loss}

% \Author[affil]{given_name}{surname}

\Author[1]{Camille}{Hankel}
\Author[1,2]{Eli}{Tziperman}
%\Author[]{}{}

\affil[1]{Department of Earth and Planetary Sciences, Harvard University, 20 Oxford St, Cambridge, MA 02138}
\affil[2]{School of Engineering and Applied Sciences, Harvard University}

%% The [] brackets identify the author with the corresponding affiliation. 1, 2, 3, etc. should be inserted.

%% If an author is deceased, please mark the respective author name(s) with a dagger, e.g. "\Author[2,$\dag$]{Anton}{Smith}", and add a further "\affil[$\dag$]{deceased, 1 July 2019}".

%% If authors contributed equally, please mark the respective author names with an asterisk, e.g. "\Author[2,*]{Anton}{Smith}" and "\Author[3,*]{Bradley}{Miller}" and add a further affiliation: "\affil[*]{These authors contributed equally to this work.}".

\correspondence{Camille Hankel (camille\_hankel@g.harvard.edu)}

\runningtitle{TEXT}

\runningauthor{TEXT}

\received{}
\pubdiscuss{} %% only important for two-stage journals
\revised{}
\accepted{}
\published{}

%% These dates will be inserted by Copernicus Publications during the typesetting process.

\firstpage{1}

\maketitle

\begin{abstract}
Abrupt and irreversible winter Arctic sea-ice loss may occur under anthropogenic warming due to the collapse of a sea-ice equilibrium at a threshold value of CO$_2$, commonly referred to as a tipping point. Previous work has been unable to conclusively identify whether a tipping point in Arctic sea ice exists because fully-coupled climate models are too computationally expensive to run to equilibrium for many CO$_2$ values. Here, we explore the deviation of sea ice from its equilibrium state under realistic rates of CO$_2$ increase to demonstrate how a few time-dependent CO$_2$ experiments can be used to predict the existence and timing of sea-ice tipping points without running the model to steady-state. This study highlights the inefficacy of using a single experiment with slow-changing CO$_2$ to discover changes in the sea-ice steady-state, and provides an alternate method that can be developed for the identification of tipping points in realistic climate models.
\end{abstract}

%\copyrightstatement{TEXT} %% This section is optional and can be used for copyright transfers.

\introduction  %% \introduction[modified heading if necessary]
The Arctic is warming at a rate at least twice as fast as the global mean with profound consequences for its sea ice cover. Summer sea ice is already exhibiting rapid retreat with warming \citep{nghiem2007rapid,stroeve2008arctic,notz2016observed}, shortening the time that socioeconomic and ecological systems have to adapt. These concerns have motivated a large body of work dedicated to both observing present-day sea ice loss \citep{kwok2011thinning,stroeve2012arctic,lindsay2015arctic,lavergne2019version} and modeling sea ice to understand whether its projected loss is modulated by a threshold-like or ``tipping point'' behavior. Abrupt loss or a tipping point in Arctic sea ice could be driven by local positive feedback mechanisms \citep{Curry-Schramm-Ebert-1995:sea, abbot2008sea,Abbot-Walker-Tziperman-2009:can,kay2012influence,Leibowicz-Abbot-Emanuel-et-al-2012:correlation,burt2016dark,feldl2020sea,hankel2021role}, remote feedback mechanisms that increase heat flux from the mid-latitudes \citep{holland2006future,park2015attribution}, or by the natural threshold corresponding to the seawater freezing point \citep{bathiany2016potential}. Such a tipping point is mathematically understood as a change in the number or stability of steady-state solutions \citep{Ghil-Childress-1987:topics,Strogatz-1994:nonlinear} as a function of CO$_2$ and is also known as a ``bifurcation". While most studies have concluded that there is no tipping point during the transition from perennial to seasonal ice cover (i.e., during the loss of \textit{summer} sea ice), the existence of a tipping point during the loss of \textit{winter} sea ice (transition to year-round ice-free conditions) continues to be debated in the literature \citep{eisenman2007arctic, Eisenman-Wettlaufer-2009:nonlinear,notz2009future, eisenman2012factors}, with three out of seven GCMs that lost their winter sea ice completely in the CMIP5 Extended RCP8.5 Scenario demonstrating an abrupt change that qualitatively looks like a tipping point, and \textit{may} be related to a bifurcation \citep{Hezel-Fichefet-Massonnet-2014:modeled,hankel2021role}. However, given the projected rapid changes to CO$_2$ in the coming centuries and the slower response of the climate system, we do not expect future sea ice to be fully equilibrated to the CO$_2$ forcing at a given time. Thus, we are interested in projecting the timing of abrupt winter Arctic sea ice changes under rapidly changing CO$_2$ forcing, when the standard steady-state tipping point analysis is not applicable.

Tipping points imply a bi-stability (meaning that sea ice can take on different values for the same CO$_2$ concentration), and hysteresis --- an irreversible loss of sea ice even if CO$_2$ is later reduced. The computational efficiency of simple models allowed studies using them to calculate the region of winter sea-ice bi-stability by running simulations to steady-state at many different CO$_2$ values, which is not possible with expensive state-of-the-art Global Climate Models (GCMs). GCM studies therefore tend to use a single experiment with very gradual CO$_2$ increases and decreases \citep{li2013transient} or even a faster CO$_2$ change \citep{ridley2012reversible, armour2011reversibility}, assuming such a run should approximate the behavior of the steady-state at different CO$_2$ concentrations. However, \cite{li2013transient} further integrated two apparently bi-stable points and found that they equilibrated to the same value of winter sea ice: there was no ``true'' bi-stability at these two CO$_2$ concentrations. This calls into question the current use of time-changing CO$_2$ runs to study the bifurcation structure of sea ice.

In light of the difficulties in using model runs with time-changing CO$_2$ (hereafter ``transient runs'') for identifying tipping points, we identify a need to understand the relationship between these transient runs and the steady-state value of sea ice as a function of CO$_2$ in systems with and without bifurcations. Theoretical work \citep{haberman1979slowly,mandel1987slow, baer1989slow, tredicce2004critical} and studies related to bi-stability in the Atlantic Meridional Overturning Circulation \citep{kim2021feedback,an2021rate} have examined tipping points when the forcing parameter (CO$_2$ in our case) changes in time at a finite rate, and found that as the forcing parameter passes the bifurcation point, the system continues to follow the old equilibrium solution for some time before it rapidly transitions to the new one. This type of analysis has to our knowledge not yet been applied in the context of winter sea ice loss under time-changing CO$_2$ concentrations, nor compared in systems with and without a bifurcation.

In order to analyze how the hysteresis curve of sea ice under time-changing forcing relates to the steady-state behavior, we run a simple physics-based model of sea ice \citep{eisenman2007arctic}, configured in three different scenarios: with a large region of bi-stability, a small region of bi-stability, and no bi-stability in the equilibrium. These three scenarios span the range of possible behaviors of winter sea ice in state-of-the-art climate models. Each case is run with different rates of CO$_2$ increase (ramping rates). We use results from this model and from an even simpler 1D dynamical system to demonstrate that the convergence of the transient behavior (under time-changing forcing) to the equilibrium behavior is very slow as a function of the ramping rate of CO$_2$. In other words, even model runs with very slow-changing CO$_2$ forcing may simulate sea ice that is considerably out of equilibrium near the period of abrupt sea ice loss. Finally, we propose an approach for uncovering the underlying equilibrium behavior in comprehensive models where it is computationally inefficient to simulate steady-state conditions for many CO$_2$ values.  

Some GCMs seem to exhibit a tipping point in winter sea ice, and others don't \citep{Hezel-Fichefet-Massonnet-2014:modeled,hankel2021role}. The reasons are likely complex and involve numerous differences in parameters and parameterizations. It is not obvious how to modify parameters in a single GCM to display all different behaviors. Therefore, we choose to use an idealized model of sea ice where we can directly produce different bifurcation behaviors to answer the question: is it possible to identify the CO$_2$ at which tipping points occur without running the model to a steady state for many CO$_2$ values? Answering such a question is an obvious prerequisite to tackling the problem of identifying climate bi-stability in noisy, high-dimensional, GCMs. In order to perform this analysis for each of the three scenarios mentioned above, we modify the strength of the albedo feedback via the choice of surface albedo parameters. The albedo values used here to generate the three scenarios are not meant to reflect realistic albedo values, but rather allow us to represent in a single model the range of sea ice equilibria behaviors that exist in different GCMs. We, therefore, follow in the footsteps of previous studies \citep[e.g.,][]{eisenman2007arctic} that have also changed parameters (the latent heat of fusion) outside of their physically relevant regime in order to understand \textit{summer} sea ice bifurcation behavior; here we follow the same approach to understand when a \textit{winter} sea ice bifurcation can be detected without running an expensive climate model to steady-state.

\section{Methods}
\subsection{Sea ice model}

The Eisenman model contains four state variables: sea ice effective thickness ($V$, which is volume divided by the area of the model grid box), sea ice area ($A$), sea ice surface temperature ($T_i$), and mixed layer temperature ($T_{ml}$) for a single box representing the entire Arctic. The atmosphere is assumed to be in radiative equilibrium with the surface, and the model is forced with a seasonal cycle of insolation, of poleward heat transport, and of local optical thickness of the atmosphere, which represents cloudiness. The full equations of the sea ice model can be found in the original paper \citep{eisenman2007arctic} and in the online Supporting Information; here, we highlight a few minor ways in which our implementation differs. First, for simplicity, we do not model leads, which in the original model were represented by capping the ice fraction at 0.95 rather than 1. Second, we use an approximation to the seasonal cycle of insolation \citep{hartmann2015global} using a latitude of 75N. The atmospheric albedo is set to 0.425 to produce the same magnitude of the seasonal cycle as in the original model of \cite{eisenman2007arctic}.

\subsection{Setup of simulations}

In our transient-forcing scenarios (described below), we vary CO$_2$ in time which affects the mid-latitude temperature ($T_{\mathrm{mid-lat}}$) and the atmospheric optical depth ($N$) (see Supporting Information). Specifically, we increase the annual mean of $T_{\mathrm{mid-lat}}$ by 3 $^{\circ}$C per CO$_2$ doubling and $N$ by a $\Delta N$ that corresponds to 3.7 W/m$^2$ per doubling. All model parameters are as in \citep{eisenman2007arctic} except as mentioned below. 

We configure the model in three different scenarios that yield a wide CO$_2$ range of bi-stability in winter sea ice (Scenario 1), a small range of bi-stability in winter sea ice (Scenario 2), and no bi-stability in winter sea ice (Scenario 3). We do so by modifying the strength of the ice-albedo feedback by changing the albedos of bare ice ($\alpha_i$), melt ponds ($\alpha_{mp}$), and ocean ($\alpha_o$), as listed in Table S1.

In each of the three scenarios, we tune the model (by adjusting the mean and amplitude of the atmospheric optical depth) to roughly match the observed seasonal cycle of ice thickness under pre-industrial CO$_2$ \citep[$\sim$ 2.5--3.7 m,][]{eisenman2007arctic}. We then run each scenario with multiple CO$_2$ ramping rates (expressed in ``years per doubling'') with an initial stabilization period (fixed pre-industrial CO$_2$), a period of exponentially increasing CO$_2$ concentration (which corresponds to linearly increasing radiative forcing), another period of stabilization at the maximum CO$_2$, a period of decreasing CO$_2$, and a final period of stabilization at the minimum CO$_2$ value (see Supplemental Figure S2).  Scenarios 2 and 3 are ramped to higher final CO$_2$ values than Scenario 1 so that they lose all their sea ice. We also directly calculate the steady-state behavior of the sea ice (as done in the original study) by running many simulations with fixed CO$_2$ values until the seasonal cycle of all the variables stabilizes. Because we expect multiple equilibria (which could be ice-free, seasonal ice, or perennial ice) at some CO$_2$ values in Scenarios 1 and 2, we run these steady-state simulations starting with both a cold (ice-covered) and a warm (ice-free) initial condition in order to find these different steady-states. In the ice-free initial condition runs, the ice-albedo feedback will still play an important role if the temperature cools sufficiently for ice to develop. At CO$_2$ values for which the sea ice is bistable, the ice-free initial condition evolves to a perennially ice-free steady-state, and the ice-covered initial condition evolves to a seasonally ice-covered steady-state (seen by the dotted and dashed lines respectively in Figs.~1a and 1c).

\subsection{Cubic ODE}

It turns out the main points we are trying to make about the transient versus equilibrium behavior of winter sea ice near a tipping point are not unique to the problem of winter sea ice, and in order to demonstrate this, we use the simplest mathematical model that can display tipping points. The cubic ODE used, while much simpler than the sea ice model above, has some of the key characteristics of the sea ice system (it is a non-autonomous system due to the time-depending forcing and has saddle-node bifurcations), which allows for direct comparison between the two models. The ODE equation,
\begin{linenomath}
\begin{align}
   \frac{dx}{dt} &= -x^3 +\delta x + \beta(t), \qquad 
   \beta(t) = \beta_0 + \mu t,
\end{align}
\end{linenomath}
contains a time-changing forcing parameter, $\beta(t)$. We consider this differential equation in three scenarios, paralleling those used with the sea ice model: in Scenario 1, $\delta=5$ leading to a wide region of bi-stability; in Scenario 2, $\delta=1$ leading to a narrow region of bi-stability, and finally, in Scenario 3, $\delta=0$ leading to a mono-stable system. The different values of $\delta$, therefore, produce the same three scenarios that were achieved in the sea ice model by modifying the strength of the ice-albedo feedback. We mimic the hysteresis experiments of the sea ice model with a sequence of ramping up and ramping down (using different ramping rates, $\mu$) with values of $\beta$ ranging from $-10$ to $10$ to sweep the parameter space that contains the bifurcations. We calculate the steady-states with fixed values of $\beta$ ($\mu=0$), starting  with both a positive and a negative initial condition of $x$ to yield two stable solutions when these exist.

We want to calculate the upper and lower CO$_2$ values of the hysteresis region in runs with time-changing (i.e., transient) CO$_2$ forcing. We do so by calculating the CO$_2$ value at which the March sea ice area drops below a critical threshold (50\% ice coverage; results are insensitive to the specific value used) during increasing and decreasing CO$_2$ integrations: we denote these CO$_2$ values $CO_2^i$ and $CO_2^d$, respectively (see Supplemental Figure S9). The difference between CO$_2^i$ and CO$_2^d$ is referred to below as the ``transient hysteresis width''; this width approaches the width of bi-stability at very slow ramping rates. 

\subsection{Predicting the CO$_2$ of the sea ice tipping point}

In order to estimate the values of CO$_2^i$ and CO$_2^d$ that would have occurred for an infinitely slow ramping rate (in other words, the range of CO$_2$ for which there is bi-stability) without having to run a model to equilibrium for all values of CO$_2$ forcing, we fit a polynomial of the form $f(x) = mx^c+b$ to CO$_2^i$ and CO$_2^d$ as functions of the ramping rate $x$. Because $c$ is negative, the fitted parameter $b$ represents the prediction of CO$_2^i$ and CO$_2^d$ at infinitely slow ramping rates, i.e., in the steady state. We also calculate the uncertainty on the fitted parameter $b$ by block-bootstrapping to account for auto-correlation; see Supporting Information. Other fits to CO$_2^i$ and CO$_2^d$ as a function of ramping rates, such as an exponential function $f(x) = a+b\exp(-cx)$ could in principle be used, although we found the fit to be less good in our case. 

\section{Results}

In the following three subsections we discuss the behavior of the sea ice model and the cubic ODE under time-changing forcing, the relationship of the transient and equilibrium behaviors, and a method that we propose for inferring the existence and location of tipping points from the transient behavior. 

\subsection{Transient response of sea ice to time-changing CO$_2$}

In Figs.~1b,d,f we plot the results of running all three scenarios (wide range of bi-stability (Scenario 1), narrow range of bi-stability (2), and no bi-stability (3)) under time-changing (transient) and fixed CO$_2$ values. In all scenarios, the experiments run with time-changing CO$_2$ exhibit transient hysteresis; the transient hysteresis width (lower horizontal gray bar in Fig.~1a) is larger for faster ramping rates (Figs.~1a,c,e). In Scenarios 1 and 2, whose equilibrium solutions (dashed and dotted black lines in Fig.~1) have a tipping point and therefore an infinite gradient of sea ice thickness vs.~CO$_2$, the faster ramping rates also lead to more gradual (and finite) gradient of sea ice thickness vs.~CO$_2$. The transient hysteresis loops across all scenarios at fast enough ramping rates (loops composed of the darkest blue and darkest red) are qualitatively similar in shape. This similarity indicates that from a single hysteresis run with time-changing CO$_2$ we cannot discern whether the underlying equilibrium behavior has a region of bi-stability or not, nor how wide the region of true bi-stability is. This result demonstrates that the apparent transient hysteresis loop found by Li et al. \cite{li2013transient} could be due to a system with or without a true hysteresis (i.e. bi-stability in the steady-state behavior), consistent with their analysis.

The robustness and generality of the above results of the sea ice model are now demonstrated by showing that the simpler ODE (eqn.~1) produces the same behavior. The 1D ODE is also configured in three scenarios with wide bi-stability (Scenario 1), narrow bi-stability (Scenario 2), and no bi-stability (Scenario 3). In Figs.~1b,d,f we see transient hysteresis in all scenarios, similar to the result from the sea ice model. Specifically, even when there is only one stable equilibrium solution in both models (Scenario 3, panels e and f), there is still a narrow region of transient hysteresis. Thus, we find that the lack of distinction in transient hysteresis loops between systems with and without bifurcations and the widening of the hysteresis loop with increased forcing parameter ramping rate appear to be robust results across these dynamical systems. Mathematically, this 1D system is fundamentally different from the sea ice model because it is not periodically forced. We show in the supplementary that adding a sinusoidal forcing term to the ODE does not qualitatively change our results.

\subsection{Slow convergence of the transient hysteresis to the equilibrium behavior}

As we saw in Fig.~1, the loss of sea ice with increasing CO$_2$ is very abrupt in the equilibrium (dashed and dotted black lines) and is infinite at the tipping point in Scenarios 1 and 2. On the other hand, the gradient is gradual and finite under time-changing forcing (blue and red curves), but steepens as the ramping rate decreases. We now quantify the rate of this steepening by examining the maximum gradient of sea ice loss during each transient simulation as a function of ramping rate (inverse of the years per doubling of CO$_2$). Our objective is to demonstrate that it is difficult to approach the equilibrium behavior using slower and slower-changing CO$_2$ runs (transient hysteresis experiments). 

In Fig.~2a, we plot the maximum gradient of March sea ice thickness \textit{with respect to CO$_2$} during each hysteresis experiment, as a function of the CO$_2$ ramping rate. In Scenarios 1 and 2 (wide and narrow bi-stability respectively), the maximum gradient gets greater as the ramping rate is slower (Fig.~2a), consistent with Fig.~1 (e.g., steepening from dark blue to light blue curves in Figs.~1a,b). In particular, it approximately follows a negative power law as a function of ramping rate on both warming and cooling time series (dashed and solid lines in Fig.~2a). In Scenario 3, the maximum gradient is nearly insensitive to the ramping rate. In Fig.~2b, we see a similar result for the simple ODE, as seen by the shallowing of the power law from Scenarios 1 to 3 (though here the slope in Scenario 3 is clearly nonzero). Notably, the power law in the case with the largest region of bi-stability (Scenario 1) is approximately given by $\max({dx}/{d\beta}) \propto \mu^{-1}$, where $\mu$ again is the ramping rate. A dependence of the maximum gradient on (ramping rate)$^{-1}$ in the case of wide bi-stability suggests that running a climate model with twice as gradual CO$_2$ ramping, leads to less than a factor of two increase in the gradient $\max(dV/d\mathrm{CO}_2)$. This is an important result because this implies that the distance between the CO$_2$ at the simulated transient ``tipping point'' and the CO$_2$ of the true (equilibrium) tipping point (which we want to estimate) only reduces by a factor of two. Thus, using more and more gradual ramping experiments may be an inefficient way to approach the equilibrium behavior of a physical system. The Supplementary Information further explains the above convergence rate of $\mu^{-1}$. 

\subsection{Predicting the steady-state behavior of sea ice using only transient runs}
One of our key results, presented next, is a method for finding the CO$_2$ concentration at which a bifurcation (if any) occurs in the equilibrium and estimating the associated hysteresis width using computationally feasible transient model runs. We are interested in this CO$_2$ concentration because it determines the threshold beyond which significant sea ice loss is practically irreversible \cite{ritchie2021overshooting}. In Fig.~3a, we plot a measure of the CO$_2$ values of the upper and lower edges of the transient hysteresis (by calculating the CO$_2$ at which the March sea ice area crosses a critical threshold, see Methods and Supplementary Figure S9). We plot this for the warming (increasing greenhouse concentration) trajectories in blue (CO$_2^i$) and for the cooling (decreasing greenhouse) trajectories in red (CO$_2^d$), as a function of the ramping rate for all three scenarios. As expected, as the ramping rate gets slower CO$_2^i$ and CO$_2^d$ asymptote to the CO$_2$ values corresponding to the edges of bi-stability and the location of the true tipping points in the case of Scenarios 1 and 2 (denoted by the $\times$ symbols). In Scenario 3, CO$_2^i$ and CO$_2^d$ asymptote to the same value (transient hysteresis width approaches zero) because there is no bi-stability in the steady-state.

Finally, we demonstrate that fitting a curve to the edges of the transient hysteresis (CO$_2^i$ and CO$_2^d$) as a function of the ramping rate can be used to predict CO$_2^i$ and CO$_2^d$ at infinitely slow ramping rates, and therefore to estimate the CO$_2$ value corresponding to a bifurcation in the equilibrium behavior without running a model to a steady-state. In Fig.~3a we plot CO$_2^i$ and CO$_2^d$, and the curves that fit them (see Methods) as functions of the ramping rate, and the predicted values of CO$_2^i$ and CO$_2^d$ at infinitely slow ramping rates with a 95\% confidence interval range shaded around them. We perform this fitting and estimation process using all the ramping experiments (18 different ramping rates total, as shown in Fig.~3a). We then repeat the fit using fewer and fewer experiments to explore how the uncertainty on predicted values of CO$_2^i$ and CO$_2^d$ increases as we move to only using a few fast ramping experiments that are more feasible when using full complexity climate models. Fig.~3b shows a summary of these analyses.

The predicted values of CO$_2^i$ and CO$_2^d$ are remarkably accurate for all scenarios (points approaching the red and blue $\times$ in Fig.~3b), even when excluding several of the slower ramping experiments. The uncertainties (indicated by the shaded blue and red bars around the points) in the predictions grow when excluding more experiments from the curve fitting process but still remain very low, especially for Scenarios 1 and 2. In predicting CO$_2^d$ for Scenario 3, the uncertainties are a bit higher because the exponential form of our fit does not represent this case as well as the others, leading to serial correlation in the residuals. Finally, we can use the difference of the distributions CO$_2^i$ and CO$_2^d$ to calculate the probability that bi-stability-- and thus a tipping point-- exists (see Supplementary Information). Overall, these results demonstrate the potential for using several shorter runs with time-changing CO$_2$ forcing to estimate the CO$_2$ value of the tipping points and predict the existence of bi-stability in GCMs where equilibrium runs or long, slow-ramping hysteresis runs are computationally infeasible.

\section{Discussion}

We have shown that it is not feasible to use a single climate model run with time-changing (transient) forcing to estimate the true location of tipping points, the range of bi-stability in the steady-state, and even the existence of bi-stability at all, consistent with the findings of \citet{li2013transient}. We also showed that this seems to be a general issue in nonlinear systems, as the same problem occurs in a generic ODE undergoing transient hysteresis. Examining the maximum gradient of sea ice thickness with respect to CO$_2$ as a function of the ramping rate of CO$_2$, we find that very long model runs are needed to identify whether this value approaches infinity, which would indicate a bifurcation, and at what CO$_2$ this occurs. Instead, we propose using a few fast-ramping experiments to predict the true range of bi-stability and provide uncertainty estimates on this prediction. The ramping rates used here likely represent an upper bound for applying our method to GCMs (for example, in the context of the abrupt transition to a moist greenhouse \citep{popp2016transition}, runaway greenhouse \citep{goldblatt2013low}, or snowball Earth state \citep{Hyde-Crowley-Baum-et-al-2000:neoproterozoic}), as we expect GCMs to have longer equilibration timescales than the idealized Eisenman sea ice model.

We demonstrated that the method we propose can accurately predict the steady-state behavior of sea ice in a simple model;  however, several challenges remain to deploying this method for use in full-complexity models. GCMs contain significant stochastic variability and multiple timescales of forcings that may render the calculated values of the diagnostics used here (such as the width of the transient hysteresis) uncertain. In addition, the functional form to fit to CO$_2^i$ and CO$_2^d$ in a GCM may require some further experimenting (such as trying an exponential rather than polynomial form) due to the more complex sea ice dynamics of the GCM. Nonetheless, we argue that using multiple runs to estimate the width of the bi-stability of a given climate variable and provide a quantified uncertainty on such a prediction offers a potential improvement over using a single hysteresis experiment. This approach still requires significant computational resources due to the need to run the model to equilibrium after the ramping up and ramping down of CO$_2$ in a hysteresis experiment.

Previous work has typically sought to identify bi-stability in sea ice because it would imply irreversibility of sea ice loss (in the sense that CO$_2$ would have to be reduced beyond the tipping point value to allow sea-ice re-growth). Here, we highlight a different perspective by focusing on realistic rates of CO$_2$ increase in addition to the steady-state behavior of sea ice. The SSP585 Scenario in CMIP6 corresponds to a ramping rate of approximately 60 years per CO$_2$ doubling: a rate at which sea ice in our idealized model already exhibits significant deviation from its steady state (60 years per doubling would fall between the 25 and 100 years per doubling blue curves in Figure 1, see also Fig.~S2). Since we identify transient hysteresis in sea ice here in all scenarios even without a deep ocean and subsequent recalcitrant warming \citep{Held-Winton-Takahashi-et-al-2010:probing}, we expect transient hysteresis to be even more pronounced in GCMs and in the real climate when such long-timescale components are included. We therefore conclude that irreversibility \textit{on policy-relevant timescales} is likely to occur in the real climate system regardless of whether an actual bifurcation (tipping point) in the equilibrium exists.

%% The following commands are for the statements about the availability of data sets and/or software code corresponding to the manuscript.
%% It is strongly recommended to make use of these sections in case data sets and/or software code have been part of your research the article is based on.

\codeavailability{An implementation of the Eisenman 2007 sea ice model in python used for this study can be found on Zenodo at: \url{https://doi.org/10.5281/zenodo.6708812} \citep{camillehankel_2022_6708812}.} %% use this section when having only software code available

%\dataavailability{TEXT} %% use this section when having only data sets available

%\codedataavailability{TEXT} %% use this section when having data sets and software code available

%\sampleavailability{TEXT} %% use this section when having geoscientific samples available

%\videosupplement{TEXT} %% use this section when having video supplements available

%\appendix
%\section{}    %% Appendix A

%\subsection{}     %% Appendix A1, A2, etc.

%\noappendix       %% use this to mark the end of the appendix section. Otherwise the figures might be numbered incorrectly (e.g. 10 instead of 1).

%% Regarding figures and tables in appendices, the following two options are possible depending on your general handling of figures and tables in the manuscript environment:

%% Option 1: If you sorted all figures and tables into the sections of the text, please also sort the appendix figures and appendix tables into the respective appendix sections.
%% They will be correctly named automatically.

%% Option 2: If you put all figures after the reference list, please insert appendix tables and figures after the normal tables and figures.
%% To rename them correctly to A1, A2, etc., please add the following commands in front of them:

%\appendixfigures  %% needs to be added in front of appendix figures

%\appendixtables   %% needs to be added in front of appendix tables

%% Please add \clearpage between each table and/or figure. Further guidelines on figures and tables can be found below.

\authorcontribution{CH and ET designed the research project and prepared the manuscript together, CH implemented the model and conducted the experiments.} %% this section is mandatory

\competinginterests{The authors declare no competing interests.} %% this section is mandatory even if you declare that no competing interests are present

\begin{acknowledgements}
The authors would like to thank Ian Eisenman for his helpful input during the project and for the guidance in using his sea ice model. ET thanks the Weizmann Institute for its hospitality during parts of this work. This work has been funded by the NSF Climate Dynamics program (joint NSF/NERC) grant AGS-1924538.
\end{acknowledgements}

%% REFERENCES

%% The reference list is compiled as follows:

% \begin{thebibliography}{}

% \bibitem[AUTHOR(YEAR)]{LABEL1}
% REFERENCE 1

% \bibitem[AUTHOR(YEAR)]{LABEL2}
% REFERENCE 2

% \end{thebibliography}

%% Since the Copernicus LaTeX package includes the BibTeX style file copernicus.bst,
%% authors experienced with BibTeX only have to include the following two lines:
%%
\bibliographystyle{copernicus}
\bibliography{ref.bib}

\begin{figure}
 \centerline{
    \includegraphics[width=.8\textwidth]{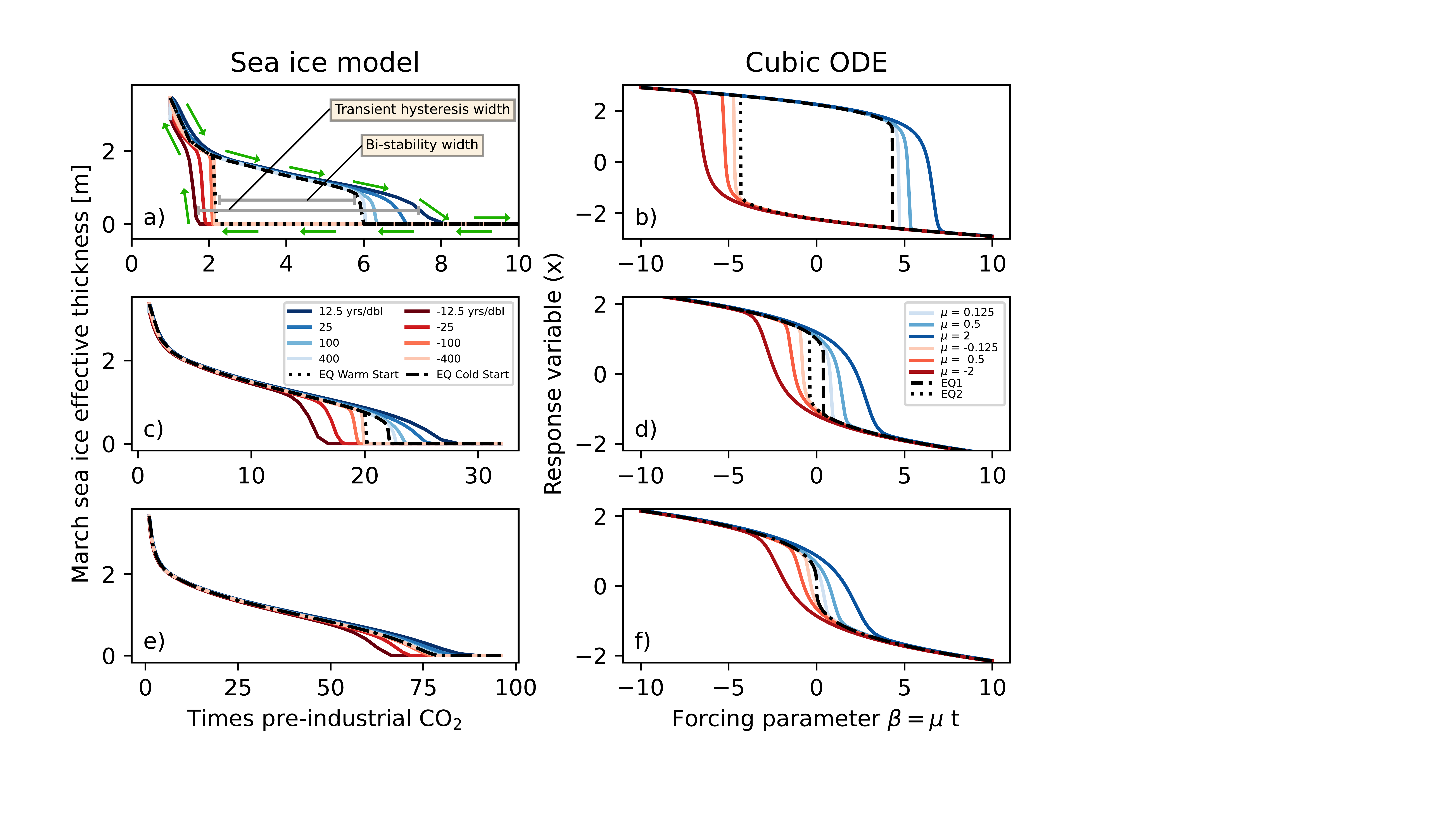}}
    \caption{Transient hysteresis runs (time-changing forcing) and equilibrium runs (fixed forcing) for average March sea ice effective thickness (sea ice volume divided by area of the grid cell; panels a,c,d) and the simple ODE from Eq.~1 (b,d,f). The first row corresponds to Scenario 1 (wide bi-stability), the second row to Scenario 2 (narrow bi-stability), and the third to Scenario 3 (no bi-stability). Blue lines indicate simulations with increasing forcing (CO$_2$ or $\beta$), while red lines indicate simulations with decreasing forcing. Dashed and dotted black lines indicate the steady-state values of sea ice or the ODE variable $x$. These two black lines are different when the two initial conditions evolve to two different steady-states. The legends indicate the different ramping rates (represented by darker colors for faster rates), which are in units of years per CO$_2$ doubling in the case of the sea ice model. The green arrows demonstrate the direction of evolving  sea ice effective thickness during the transient hysteresis experiments.}
\end{figure}

\begin{figure}
\centerline{
    \includegraphics[width=.8\textwidth]{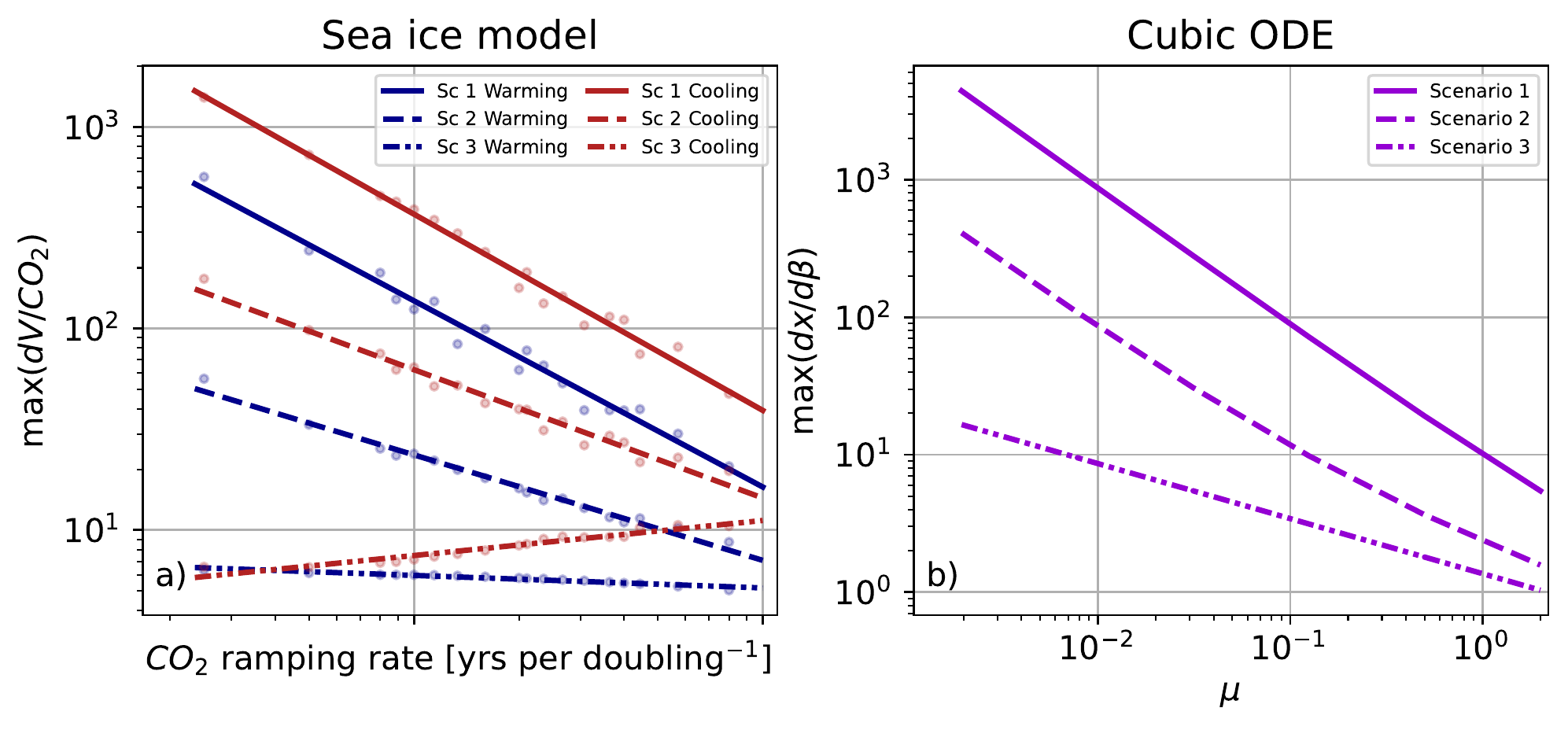}}
    \caption{Maximum gradient of sea ice effective thickness with respect to CO$_2$ in panel a, and the maximum gradient of $x$ with respect to the forcing parameter $\beta$ in panel b during transient simulations. For the sea ice model (a) the data points from the 18 different runs are shown as faded points, with a superimposed line of best fit. For the cubic ODE (b) the maximum gradient lines corresponding to increasing and decreasing forcing time series are identical due to the symmetry around $\beta =0$ seen in Fig.~1b, d, and f.}
\end{figure}

\begin{figure}
\centerline{
    \includegraphics[width=\textwidth]{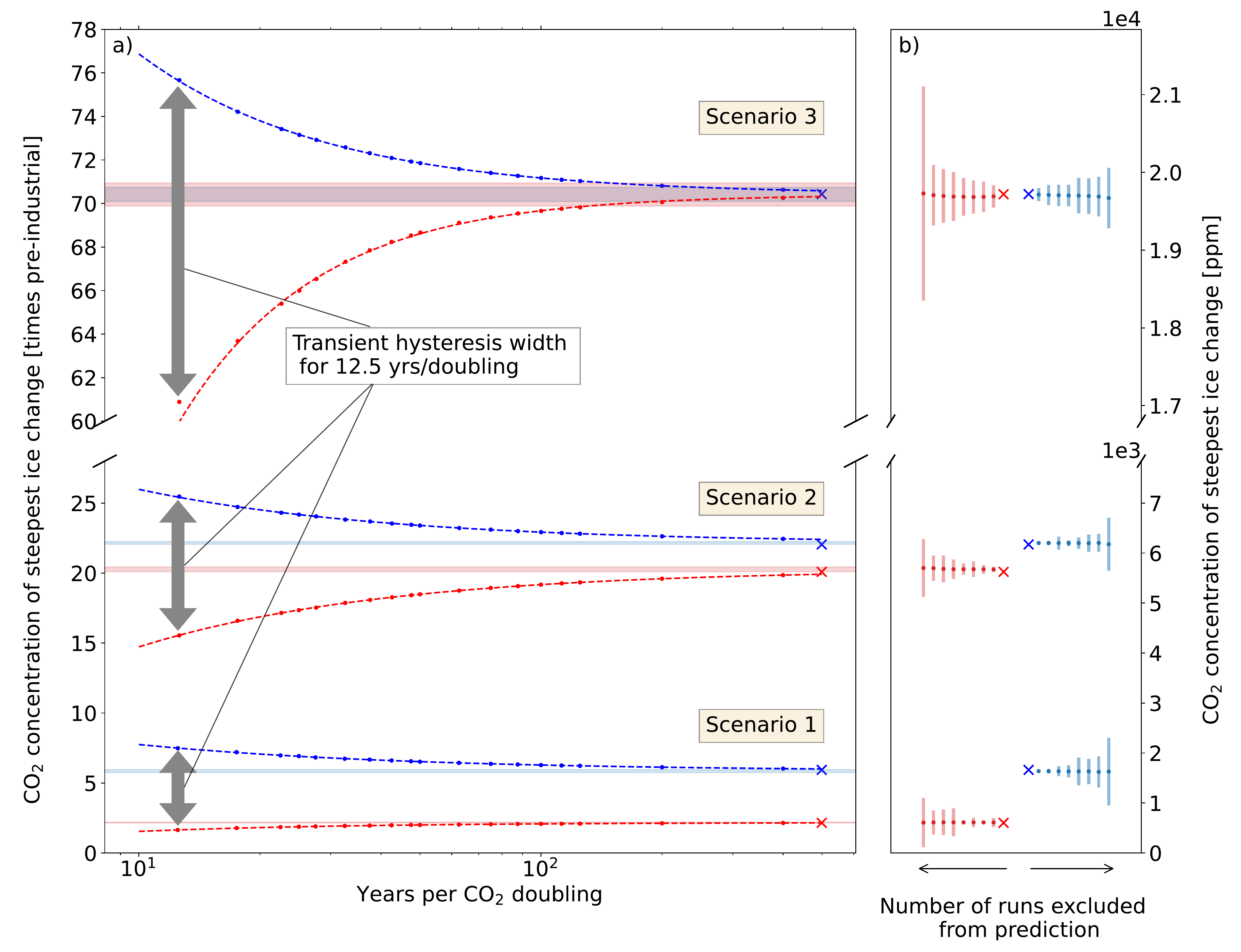}} 
    \caption{Estimating the equilibrium tipping point value from the transient hysteresis runs. In panel a, the scatter points show the CO$_2$ value of the right and left edges of the transient hysteresis (CO$_2^i$ and CO$_2^d$, located along increasing (blue) and decreasing (red) CO$_2$ time-series respectively) for different ramping rates. The dashed lines show the curve that is fitted to the scatter points, and the shaded blue and red bands show $\pm2\sigma$ around the predicted values of CO$_2^i$ and CO$_2^d$ at infinitely slow ramping rates. The blue and red $\times$'s show the true equilibrium values of CO$_2^i$ and CO$_2^d$ (calculated from the fixed CO$_2$ runs starting with cold and warm initial conditions respectively). In panel b, we analyze the accuracy of this prediction as we use fewer transient runs. For the three scenarios, we show the result of sequentially excluding the most gradual ramping simulations from the curve-fitting process used for predictions. The dots and the corresponding bars represent the predicted equilibrium values of CO$_2^i$ and CO$_2^d$, and $\pm2\sigma$ around the prediction, and dots moving away from the true value with larger error bars correspond to excluding more and more runs from the calculation.}
\end{figure}

\end{document}

\title{Supporting Information for ``An approach for projecting the timing of abrupt winter Arctic sea ice loss"}
%
% e.g., \title{Supporting Information for "Terrestrial ring current:
% Origin, formation, and decay $\alpha\beta\Gamma\Delta$"}
%
%DOI: 10.1002/%insert paper number here%

%% ------------------------------------------------------------------------ %%
%
%  AUTHORS AND AFFILIATIONS
%
%% ------------------------------------------------------------------------ %%

% List authors by first name or initial followed by last name and
% separated by commas. Use \affil{} to number affiliations, and
% \thanks{} for author notes.
% Additional author notes should be indicated with \thanks{} (for
% example, for current addresses).

% Example: \authors{A. B. Author\affil{1}\thanks{Current address, Antartica}, B. C. Author\affil{2,3}, and D. E.
% Author\affil{3,4}\thanks{Also funded by Monsanto.}}

\authors{Camille Hankel\affil{1}, Eli Tziperman\affil{1,2}}

\affiliation{1}{Department of Earth and Planetary Sciences, Harvard University, 20 Oxford St, Cambridge, MA 02138}
\affiliation{2}{School of Engineering and Applied Sciences, Harvard University}

%% ------------------------------------------------------------------------ %%
%
%  BEGIN ARTICLE
%
%% ------------------------------------------------------------------------ %%

% The body of the article must start with a \begin{article} command
%
% \end{article} must follow the references section, before the figures
%  and tables.

\begin{article}
\baselineskip=16pt\renewcommand{\baselinestretch}{0.5}

%% ------------------------------------------------------------------------ %%
%
%  TEXT
%
%% ------------------------------------------------------------------------ %%

\noindent\textbf{Contents of this file}
%%%Remove or add items as needed%%%
\begin{enumerate}
\item Text S1 to S6
\item Table S1 
\item Figures S1 to S9
%if Tables are larger than 1 page, upload as separate excel file
\end{enumerate}
\hfill \break
\noindent\textbf{Introduction}

This document contains the equations of the sea ice model used (Text S1), results from a modified version of the cubic ODE that has a periodic forcing term (Text S2), specific details of the calculations of the diagnostics on the time-changing forcing trajectories (Text S3), a heuristic derivation that explains the result that $\max (dx/d\beta) \approx 1/\mu$ (Text S4), and the calculations of the uncertainties on the predictions of bi-stability width (Text S5). Text S6 provides a method for estimating the likelihood of the existence of bi-stability using results from the main text. In Table S1 we provide the exact parameter values used to configure the sea ice model in the three scenarios described in the main text. Figures S1--S2, we provide extra information on model setup and experimental design, and Figures S3--S5 show March average quantities for all four state variables in the sea ice model from the experiments performed in the main text. Figure S6 shows results from the periodically-forced ODE described in Text S2, and Figures S7--S8 relate to understanding the convergence behavior of $\max (dx/d\beta)$ as a function of ramping rate. Finally, Figure S9 helps visualize our method for calculating CO$_2^i$ and CO$_2^d$, the two edges of transient hysteresis described in the main text.
\hfill \break

\noindent\textbf{Text S1: Sea ice model equations}

The Eisenman model contains four state variables: sea ice volume ($V$), sea ice area ($A$), sea ice surface temperature ($T_i$), and mixed layer temperature ($T_{ml}$) for a single box representing the entire Arctic. The atmosphere is assumed to be in radiative equilibrium with the surface, and the model is forced by a seasonal cycle of insolation, of poleward heat transport, and of local optical thickness of the atmosphere, which represents cloudiness. The addition of CO$_2$ is represented by increasing the optical thickness and the midlatitude temperature, which increases poleward heat flux. Melt ponds are parameterized by allowing the ice to melt when the surface temperature reaches 0 $^{\circ}$C and by modifying the ice albedo when this condition is met. The equations for the model are written below, and can also be found in the original paper \cite{eisenman2007arctic} that used them. The surface longwave radiation imbalance at the surface is
\begin{linenomath}
\begin{align}
    \epsilon(T,T_s) = \frac{2a}{2+N} - \frac{D(T_s)}{2} + b\left(T-T_s+\frac{2T_s}{2+N}\right),
\end{align}
\end{linenomath}
where $N$ is the optical depth of the atmosphere and $T_s = AT_i+(1-A)T_{ml}$ is the surface temperature of the box. $D(T_s)$ is the atmospheric poleward heat transport given by 
\begin{linenomath}
\begin{align}
    D(T_s) = k_D(T_{\mathrm{mid-lat}}-T_s). 
\end{align}
\end{linenomath}
The net heat flux into the mixed layer is given by
\begin{linenomath}
\begin{align}
    F_{ml} = (1-A)(-\epsilon(T_{ml},T_s)+(1-\alpha_o)F_{sw})-A\gamma T_{ml}+F_{entr},
\end{align}
\end{linenomath}
where $F_{sw}$ is the shortwave radiation reaching the surface. The mixed layer temperature normally evolves according to this net heat flux, except for when it is at 0$^{\circ}$ C and cooling, at which point the negative heat flux goes entirely into new ice production and $T_{ml}$ stays at 0$^{\circ}$ C. This is expressed as:
\begin{linenomath}
\begin{align}
    c_{ml}H_{ml}\frac{dT_{ml}}{dt}&= 
    \begin{cases}
    0 & \text{if } T_{ml} = 0 \text{ and } F_{ml}<0,
    \\
    F_{ml} & \text{otherwise},
    \end{cases}\\
    F_{ni}&= 
    \begin{cases}
    -F_{ml} & \text{if } T_{ml} = 0 \text{ and } F_{ml}<0,\\
    0 & \text{otherwise},
    \end{cases}
\end{align}
\end{linenomath}
where $F_{ni}$ is the new ice production. The ice volume and surface temperature evolution are conditioned on whether or not the surface is melting. With the net ice surface heat flux when $T_i=0$ written as $F_{net} = -\epsilon(0,T_s)+(1-\alpha_i)F_{sw}$, the equations for ice volume and surface temperature are: 
\begin{linenomath}
\begin{align}
    L\frac{dV}{dt} &= \begin{cases}
    A(\epsilon(T_i,T_s)-(1-\alpha_{mp})F_{sw}-\gamma T_{ml})-v_0A & T_i=0 \text{ and } F_{net}>0,\\
    A(-\frac{kT_i}{h}-\gamma T_{ml})+F_{ni}-v_0LV & \text{otherwise},
    \end{cases}
\\
    \frac{ch}{2}\frac{dT_i}{dt} &= \begin{cases}
    0 & \text{if } T_i=0 \text{ and } F_{net}>0, \\
    -\epsilon(T_i,T_s)+(1-\alpha_i)F_{sw} -\frac{kT_i}{h} & \text{otherwise}.
    \end{cases}
\end{align}
\end{linenomath}
Finally, ice area evolution occurs according to: 
\begin{linenomath}
\begin{align}
    \frac{dA}{dt} = \frac{F_{ni}}{Lh_0}-\frac{A}{2V}\mathcal{R}\left(-\frac{dV}{dt}\right)-v_0A.
\end{align}
\end{linenomath}

As mentioned in the main text, in our implementation of the model we also allow the CO$_2$ concentration to vary inter-annually, by allowing the optical depth ($N$) and the mid-latitude temperature to be functions of time. They can be written as: 
\begin{linenomath}
\begin{align}
    N(t) &= N_0 +A\sin\left(\frac{2\pi}{1 \text{ yr}}t\right) +\Delta N\times\log_2(\mathrm{CO}_2(t)/280\text{ ppm}),\\
    T_{mid-lat}(t) &=T_0 +B \sin\left(\frac{2\pi}{1 \text{ yr}}t\right) + 3\degree\text{C}\times\log_2(\mathrm{CO}_2(t)/280\text{ ppm}),
\end{align}
\end{linenomath}
where the time dependence of the final term in each equation is a modification from \cite{eisenman2007arctic}. The $\sin$ terms in each equation represent the seasonal cycles of the atmospheric optical depth and the mid-latitude temperature, respectively, at 280 ppm of CO$_2$. In our hysteresis experiments, $CO_2(t)$ is an exponentially increasing and then exponentially decreasing function of time (see Fig.~S2), leading to CO$_2$ forcings that change linearly in time.
\hfill \break

\noindent\textbf{Test S2: Cubic ODE with periodic forcing}

In this section, we analyze an ODE that is similar to the one presented in the main text but includes a periodic forcing term, which makes it more analogous to the seasonally forced model of sea ice. The equations for this system are:
\begin{linenomath}
\begin{align}
   \frac{dx}{dt} &= -x^3 +\delta x + 50\sin(2\pi t) + \beta(t), \qquad 
   \beta(t) = \beta_0 + \mu t,
\end{align}
\end{linenomath}
The magnitude of 50 on the $\sin$ term is chosen such that the magnitude of the changes in $\beta$ compared to the amplitude of the periodic forcing is roughly similar to the magnitude of CO$_2$ changes compared to the amplitude of the seasonal cycle of insolation in the sea ice model. The values of $\delta$ needed to configure the three scenarios are slightly different than those for the non-periodic ODE and are as follows: $\delta =6$ for Scenario 1 (wide bi-stability), $\delta = 4$ for Scenario 2 (narrow bi-stability) and $\delta=3.4$ for Scenario 3 (no bi-stability). Since the solution $x$ is now oscillatory during time-changing and fixed forcing scenarios, we plot the maximum values of $x$ during each oscillation; this is meant to parallel the plotting of March sea ice (which is approximately the maximum amount of ice during the annual cycle). We only range $\beta$ from -5 to 5 as this is the range that is needed to sweep across the bifurcations. We see in Figure S6 that the qualitative characteristics of the transient hysteresis found in an ODE without periodic forcing (main text) also are found in this ODE. There is transient hysteresis for all three scenarios (panels a-c) and the width of this hysteresis gets wider as we move to faster ramping rates. The addition of the periodic forcing combined with the choice to plot the maximum value of $x$ during each oscillation also generates asymmetry in the increasing and decreasing forcing trajectories (blue vs. red lines). In panel d we plot the maximum gradient of $x$ with respect to $\beta$ during transient forcing simulations versus the ramping rate, $\mu$. We see that, similar to the result in the main text, the maximum gradient follows a negative power law as a function of $\mu$, with the slope of the power law becoming steeper and approaching a value of -1 as we move from Scenario 3 to Scenario 1. Thus we conclude that the comparisons we made in the main text between the sea ice model and the simple cubic ODE would also apply if we had chosen to include a periodic forcing in the cubic ODE. 
\hfill \break

\noindent\textbf{Text S3: Calculating diagnostics on time-changing forcing trajectories}

We calculate the maximum rate of change of sea ice volume with respect to CO$_2$ concentration by taking the maximum change in monthly-averaged March sea ice volume between any two consecutive years during the transient hysteresis period of the simulations (i.e., ignoring the initial fast decline of sea ice at low CO$_2$) divided by the change in yearly average CO$_2$ concentration between those two years. The maximum rate of change of sea ice in time is calculated analogously but divided by the time interval rather than the change in yearly mean CO$_2$.

To calculate the maximum rate of change of the solution to the cubic ODE with respect to the forcing parameter ($\beta(t)$) or time, we calculate the smoothed absolute change around the two time steps that show the greatest absolute change in $x$, divided by the change in $\beta$ or by time. The ``smoothed'' absolute change is simply the difference between the mean value of $x$ over the five time steps before the largest jump in $x$ and the mean during the five time steps after the jump. 
\hfill \break

\noindent\textbf{Text S4: Deriving $\text{max}(dx/d\beta)\propto \mu^{-1}$}

To understand why $\max({dx}/{d\beta}) \propto \mu^{-1}$ and thus why the maximum rate of change of sea ice also follows a similar negative power law as a function of ramping rate, we first note that,
\begin{linenomath}
\begin{align}
    \text{max}\left(\frac{dx}{d\beta}\right) = \text{max}\left(\frac{dx}{dt}\frac{dt}{d\beta}\right) = \frac{1}{\mu}\text{max}\left(\frac{dx}{dt}\right).
\end{align}
\end{linenomath}
Thus in Fig.~S7 we plot the maximum March $dV/dt$ and the maximum $dx/dt$ respectively as a function of the ramping rate. We can see that as predicted by eqn.~12, the slopes of the power laws in main text Figs.~2a and 2b are those found in Figs.~S7 minus 1. In particular, the $\mu^{-1}$ rate of convergence in Scenario 1 is recovered when noting that $\max(dx/dt)$ appears to be a constant value as a function of the ramping rate for small enough rates; an unintuitive result that is explained further below. The slopes of the power laws that characterize the convergence of the transient simulations to their equilibrium behavior may also prove a useful tool for inferring the equilibrium, in addition to the method we proposed in the main text.

Next, we provide a heuristic derivation for why $\max(dx/dt)$ approaches a constant in Scenarios 1 and 2 when $\mu$ is small. Using Figure S8 for reference, we can see that as $x$ moves from one equilibrium, $x^*_a$, to the next, $x^*_b$, during the bifurcation, its maximum rate of change is given by the local minimum of the $dx/dt$ curve, given by $c$. However, because we are considering a non-autonomous equation with the time-changing forcing $\beta(t)$ that shifts the $dx/dt$ curve down in time, the local minimum $c$ is a function of time, according to the ramp rate, $\mu$. In other words, the maximum rate of change ($dx/dt$) during the transition from $x_a^*$ to $x_b^*$ (the value of which is also changing in time) under a given forcing $\beta(t) = \beta_0 + \mu t$ is greater than or equal to $c_0$ and less than or equal to $c=c_0 + \mu t$, where $c_0$ is the local minimum of $dx/dt$ exactly at the point of bifurcation and $t$ is the time it takes to complete the transition from one steady-state to the other. In the case where $c_0$ is large compared to $\mu t$, we can make the approximation that $c \approx c_0$ as $\mu \xrightarrow{} 0$. Thus, the $\text{max}({dx}/{dt}) \xrightarrow {}c_0$ as $\mu \xrightarrow{}0$. Finally, returning to equation 11, we get that $\max({dx}/{d\beta}) = {c_0}/{\mu}$ for small $\mu$, recovering the $\mu^{-1}$ rate of convergence we estimated empirically. 

We do not expect this derivation to hold in cases where $c_0$ is not large compared to $\mu$. Indeed, in the ODE without a bifurcation where $c_0 = 0$, we see that $\text{max}(\frac{dx}{dt})$ is a positive power of $\mu$, which, when divided by $\mu$ according to equation 11, causes $\text{max}(\frac{dx}{d\beta})$ to be a negative power of $\mu$ with a magnitude less than 1. We argue that this derivation from a simple ODE provides intuition for the more gradual slopes for the physics-based sea ice model seen in Fig.~2a as we move from a scenario (1) with a wide region of bi-stability to a scenario (3) with no bi-stability or bifurcation. Specifically, in a cubic ODE, the value $c_0$ exactly corresponds to the width of parameter forcing for which there is bi-stability; while this may not hold exactly for the physical sea ice model, we expect Scenario 1 in the sea ice model to be associated with a large $c_0$ (a fast maximum rate of change of sea ice in time), Scenario 2 to be associated with a smaller $c_0$, and Scenario 3 to be associated with small or zero $c_0$. As discussed previously, the larger the magnitude of $c_0$ the closer the slope of the maximum rate of change in time versus the ramp rate is to zero, which in turn sets the slope of the maximum rate of change in CO$_2$ versus the ramp rate. Thus, the derivation that $\max({dx}/{d\beta}) \approx c_0\mu^{-1}$ for the cubic ODE with a bifurcation provides insight into the convergence behavior of the transient sea ice simulations for all three scenarios. 
\hfill \break

\noindent\textbf{Text S5: Calculating uncertainty on predictions of the CO$_2$ value of tipping points}

When fitting curves to CO$_2^i$ and CO$_2^d$ in order to predict CO$_2^i$ and CO$_2^d$ at infinitely slow ramping rates, we noticed that there was some auto-correlation in the residuals (which are the difference between the fitted curves and the actual values of CO$_2^i$ and CO$_2^d$ at all 18 ramping rates). This means that using the covariance matrix of the fitted parameters underestimates the uncertainty on the prediction of CO$_2^i$ and CO$_2^d$ at infinitely slow ramping rates (especially in the case of Scenario 3). To address this issue, we instead use a block-bootstrapping method to calculate the uncertainty on our predictions. We sample with a block size of three, and bootstrap 1000 times, giving us a distribution of estimates of the equilibrium values of CO$_2^i$ and CO$_2^d$. From these distributions, we can calculate the standard deviation of the predictions, and we use these standard deviations to plot 95\% confidence intervals around the predictions in main text Fig.~3. We perform this block-bootstrapping procedure for each of the predictions that use fewer and fewer simulations to produce all the confidence intervals plotted in Fig.~3b. 
\hfill \break

\noindent\textbf{Text S6: Calculating the probability of the existence of bi-stability}

The block-bootstrapping process described in Text S5 gives us distributions of estimated values of CO$_2^i$ and CO$_2^d$ at infinitely slow ramping rates. Since the width of the ``true" hysteresis is the difference between CO$_2^i$ and CO$_2^d$, we can take the difference of these distributions to estimate the likelihood that the hysteresis width is greater than zero and thus that bi-stability and a tipping point exist. Using only experiments with a ramping rate of 75 years/doubling or faster, in both Scenarios 1 or 2 we find that $>$95\% of the difference distribution (hysteresis width distribution) is greater than zero; in other words, we can say that there is less than a 5\% chance that bi-stability does \textit{not} exist. Excluding even more of the ramping experiments (which would be computationally expensive in a GCM) to use only experiments with a ramping rate of 47.5 years/doubling or less, we find that there is an $>$80\% chance that bi-stability exists for both Scenarios 1 and 2. For Scenario 3 (which we know does not have bi-stability), no matter how many experiments we exclude there is never more than 55\% chance that bi-stability exists; in fact, when using few experiments the distribution skews towards predicting a \textit{negative} width of hysteresis, an unphysical result that may itself suggest the lack of bi-stability. These statistical tests can be used to estimate the binary existence or non-existence of tipping points, in addition to the method for predicting of the CO$_2$ location of tipping points presented in the main text. 

%Repeat for any additional Supporting audio files

%%% End of body of article:
%%%%%%%%%%%%%%%%%%%%%%%%%%%%%%%%%%%%%%%%%%%%%%%%%%%%%%%%%%%%%%%%
%
% Optional Notation section goes here
%
% Notation -- End each entry with a period.
% \begin{notation}
% Term & definition.\\
% Second term & second definition.\\
% \end{notation}
%%%%%%%%%%%%%%%%%%%%%%%%%%%%%%%%%%%%%%%%%%%%%%%%%%%%%%%%%%%%%%%%

%% ------------------------------------------------------------------------ %%
%%  REFERENCE LIST AND TEXT CITATIONS

%%%%%%%%%%%%%%%%%%%%%%%%%%%%%%%%%%%%%%%%%%%%%%%
% 
%
\bibliography{ref}
%
% no need to specify bibliographystyle
%
% Note that ALL references in this supporting information file must also be referenced in the primary manuscript
%
%%%%%%%%%%%%%%%%%%%%%%%%%%%%%%%%%%%%%%%%%%%%%%%
% if you get an error about newblock being undefined, uncomment this line:
%\newcommand{\newblock}{}

% \bibliography{ uncomment this line and enter the name of your bibtex file here } 

%Reference citation instructions and examples:
%
% Please use ONLY \cite and \citeA for reference citations.
% \cite for parenthetical references
% ...as shown in recent studies (Simpson et al., 2019)
% \citeA for in-text citations
% ...Simpson et al (2019) have shown...
% DO NOT use other cite commands (e.g., \citet, \citep, \citeyear, \nocite, \citealp, etc.).
%
%
%...as shown by \citeA{jskilby}.
%...as shown by \citeA{lewin76}, \citeA{carson86}, \citeA{bartoldy02}, and \citeA{rinaldi03}.
%...has been shown \cite<e.g.,>{jskilbye}.
%...has been shown \cite{lewin76,carson86,bartoldy02,rinaldi03}.
%...has been shown \cite{lewin76,carson86,bartoldy02,rinaldi03}.
%
% apacite uses < > for prenotes, not [ ]
% DO NOT use other cite commands (e.g., \citet, \citep, \citeyear, \nocite, \citealp, etc.).
%

%% ------------------------------------------------------------------------ %%
%
%  END ARTICLE
%
%% ------------------------------------------------------------------------ %%
\end{article}

\clearpage
\baselineskip=12pt\renewcommand{\baselinestretch}{0.5}

% Copy/paste for multiples of each file type as needed.

% enter figures and tables below here: %%%%%%%
%
%
%
%
% EXAMPLE FIGURES
% ---------------
% If you get an error about an unknown bounding box, try specifying the width and height of the figure with the natwidth and natheight options.
% \begin{figure}
%\setfigurenum{S1} %%You can change number for each figure if you want, not required. "S" prepended automatically.
% \noindent\includegraphics[natwidth=800px,natheight=600px]{samplefigure.eps}
%\caption{caption}
%\label{epsfiguresample}
%\end{figure}
%
%
% Giving latex a width will help it to scale the figure properly. A simple trick is to use \textwidth. Try this if large figures run off the side of the page.
% \begin{figure}
% \noindent\includegraphics[width=\textwidth]{anothersample.png}
%\caption{caption}
%\label{pngfiguresample}
%\end{figure}
%
%
%\begin{figure}
%\noindent\includegraphics[width=\textwidth]{athirdsample.pdf}
%\caption{A pdf test figure}
%\label{pdffiguresample}
%\end{figure}
%
% PDFLatex does not seem to be able to process EPS figures. You may want to try the epstopdf package.
%
%
% ---------------
% EXAMPLE TABLE
%
%\begin{table}
%\settablenum{S1} %%Change number for each table
%\caption{Time of the Transition Between Phase 1 and Phase 2\tablenotemark{a}}
%\centering
%\begin{tabular}{l c}
%\hline
% Run  & Time (min)  \\
%\hline
%  $l1$  & 260   \\
%  $l2$  & 300   \\
%  $l3$  & 340   \\
%  $h1$  & 270   \\
%  $h2$  & 250   \\
%  $h3$  & 380   \\
%  $r1$  & 370   \\
%  $r2$  & 390   \\
%\hline
%\end{tabular}
%\tablenotetext{a}{Footnote text here.}
%\end{table}
\begin{table}
%\centering
\begin{center}
\begin{tabular}{|c||c|c|c|c|c|} 
 \hline
  & $\alpha_i$ & $\alpha_{mp}$ & $\alpha_o$  & Equilibrium behavior  \\ 
 \hline
 Scenario 1 & .75 &  .45 & .1 & wide bi-stability\\ 
 \hline
 Scenario 2 & .5 & .4 & .2 & narrow bi-stability\\ 
 \hline
  Scenario 3 & .4 & .4 & .4 & no bi-stability\\ 
 \hline
\end{tabular}
\end{center}
\caption{\baselineskip=14pt
Model configurations leading to wide, narrow, and no bi-stability regimes. The symbols $\alpha_i$, $\alpha_{mp}$, and $\alpha_o$, refer to the albedo of bare ice, melt ponds, and open ocean respectively.}
\label{tab:bistability-regimes}
\end{table}

\begin{figure}
\centerline{
  \includegraphics[width=.75\textwidth]{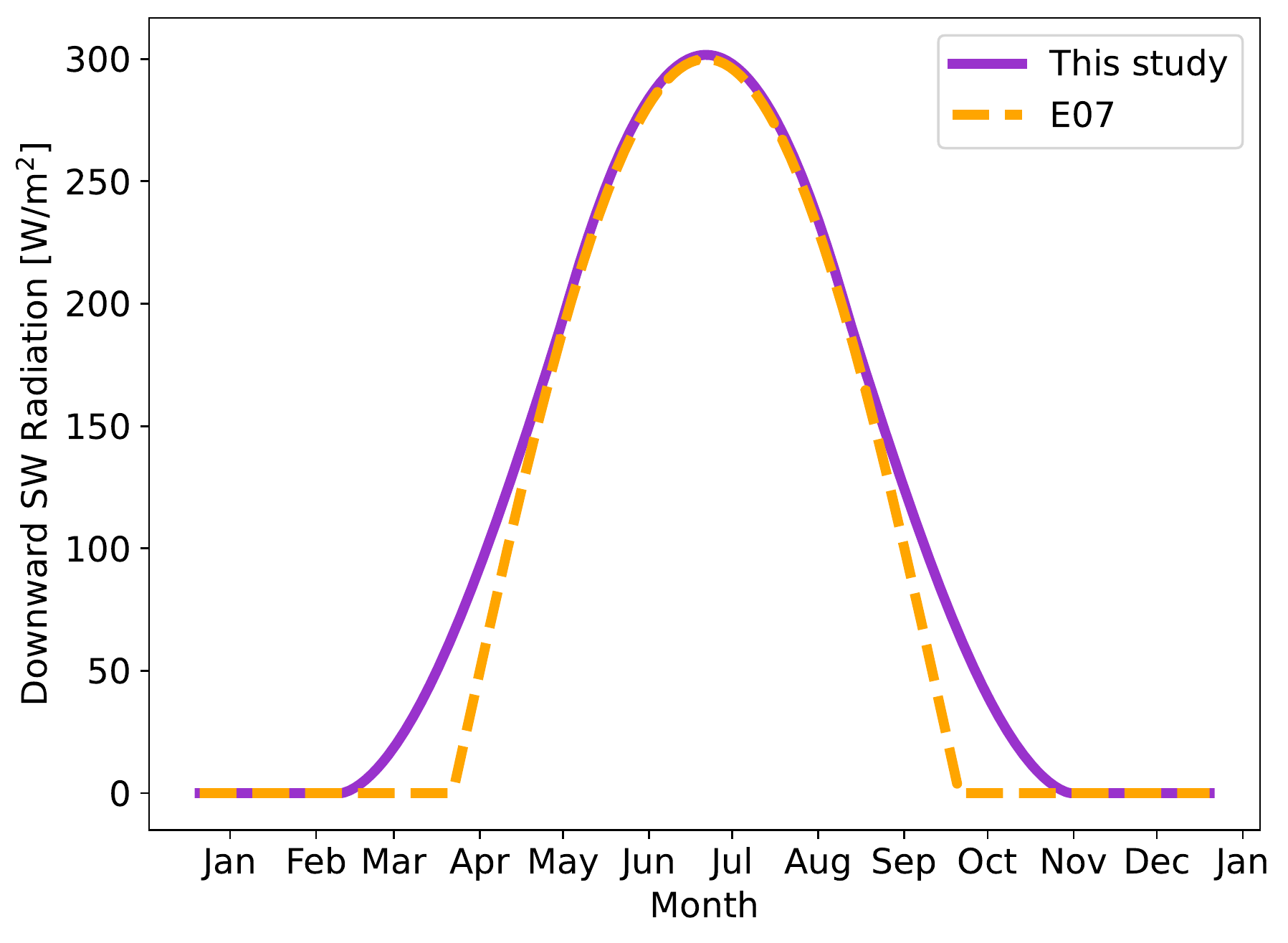}}
    \caption{\baselineskip=14pt A comparison of our seasonal cycle of insolation to that of Eisenman (2007).}
\end{figure}

\begin{figure}
\centerline{
    \includegraphics[width=.75\textwidth]{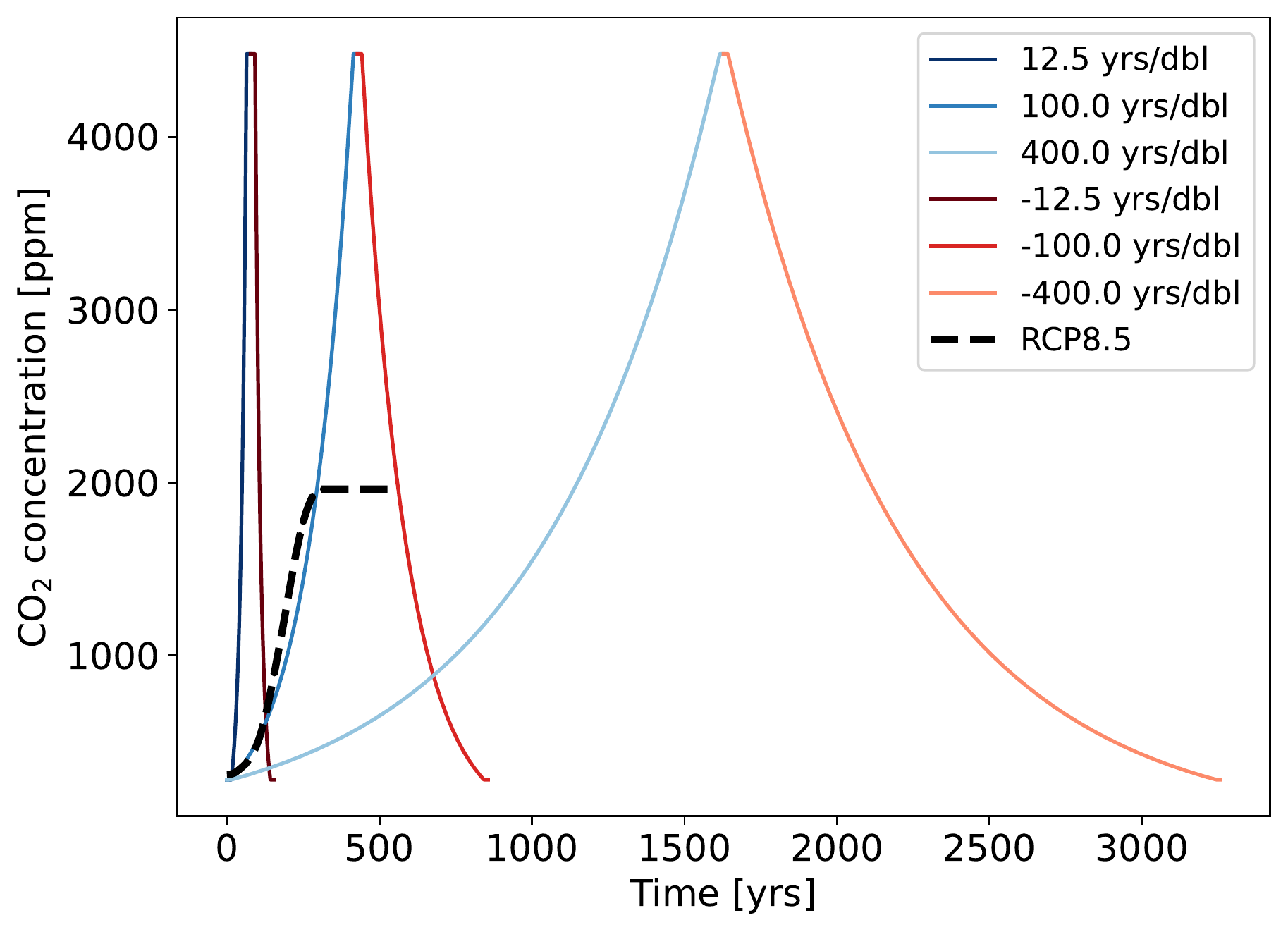}}

  \caption{
    \baselineskip=14pt
    A subset our CO$_2$ hysteresis experiments (ramp up, hold CO$_2$ fixed, ramp down) compared to the ramping rate of the RCP8.5 Scenario in CMIP5 (which is nearly the same as the ramping rate of SSP585 in CMIP6). Exponential increases in the concentration of CO$_2$ in time lead to linear increases of the CO$_2$ radiative forcing.}
\end{figure}

\begin{figure}
\centerline{
    \includegraphics[width=\textwidth]{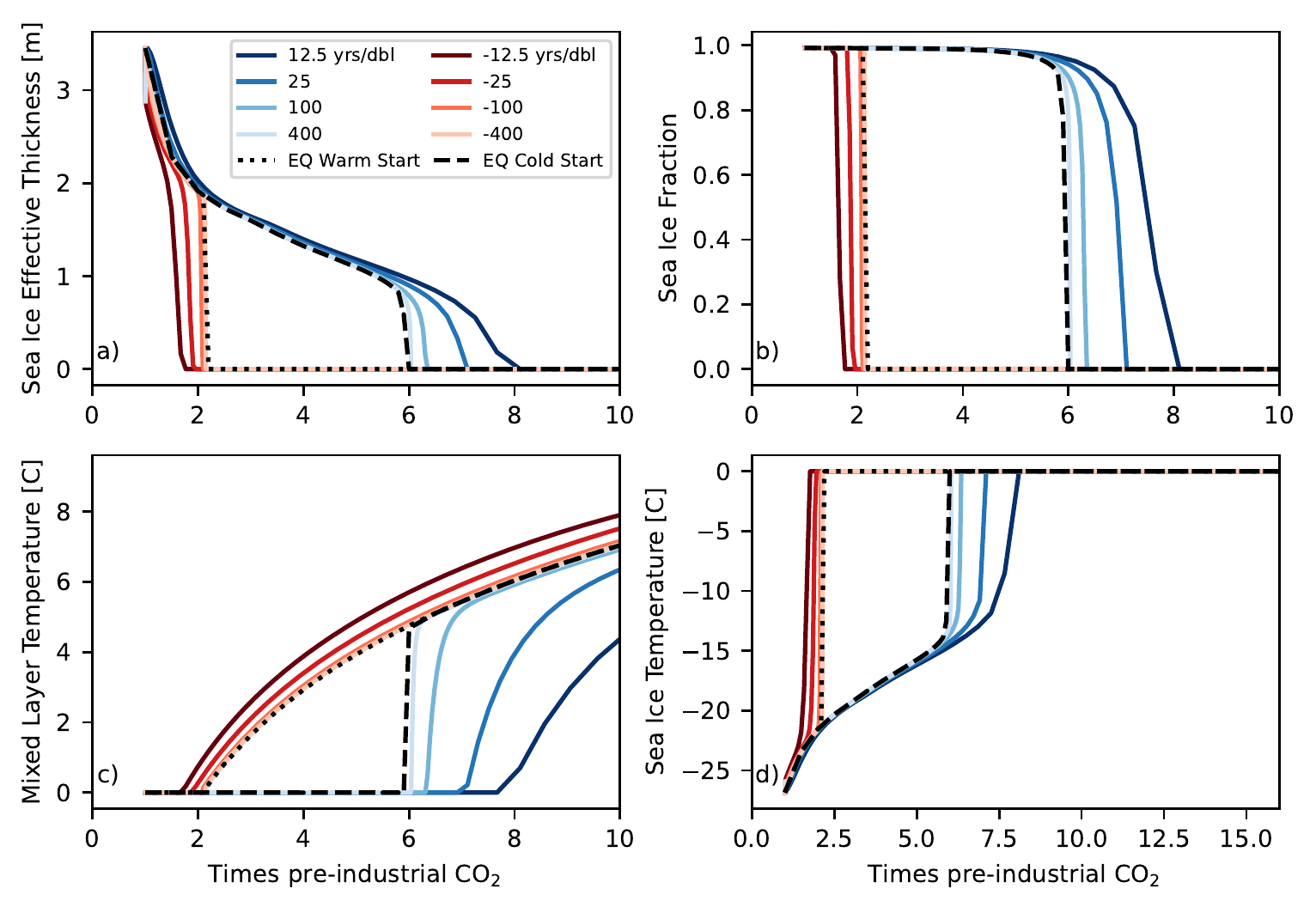}}
    \caption{\baselineskip=14pt All four state variables of the sea ice model from Scenario 1 (wide bi-stability) runs, shown as their March monthly averages: sea ice effective thickness (a), sea ice fraction (b), mixed layer temperature (c), and sea ice temperature (d). Legend and coloring are the same as in Fig.~1 in the main text.}
\end{figure}

\begin{figure}
\centerline{
    \includegraphics[width=\textwidth]{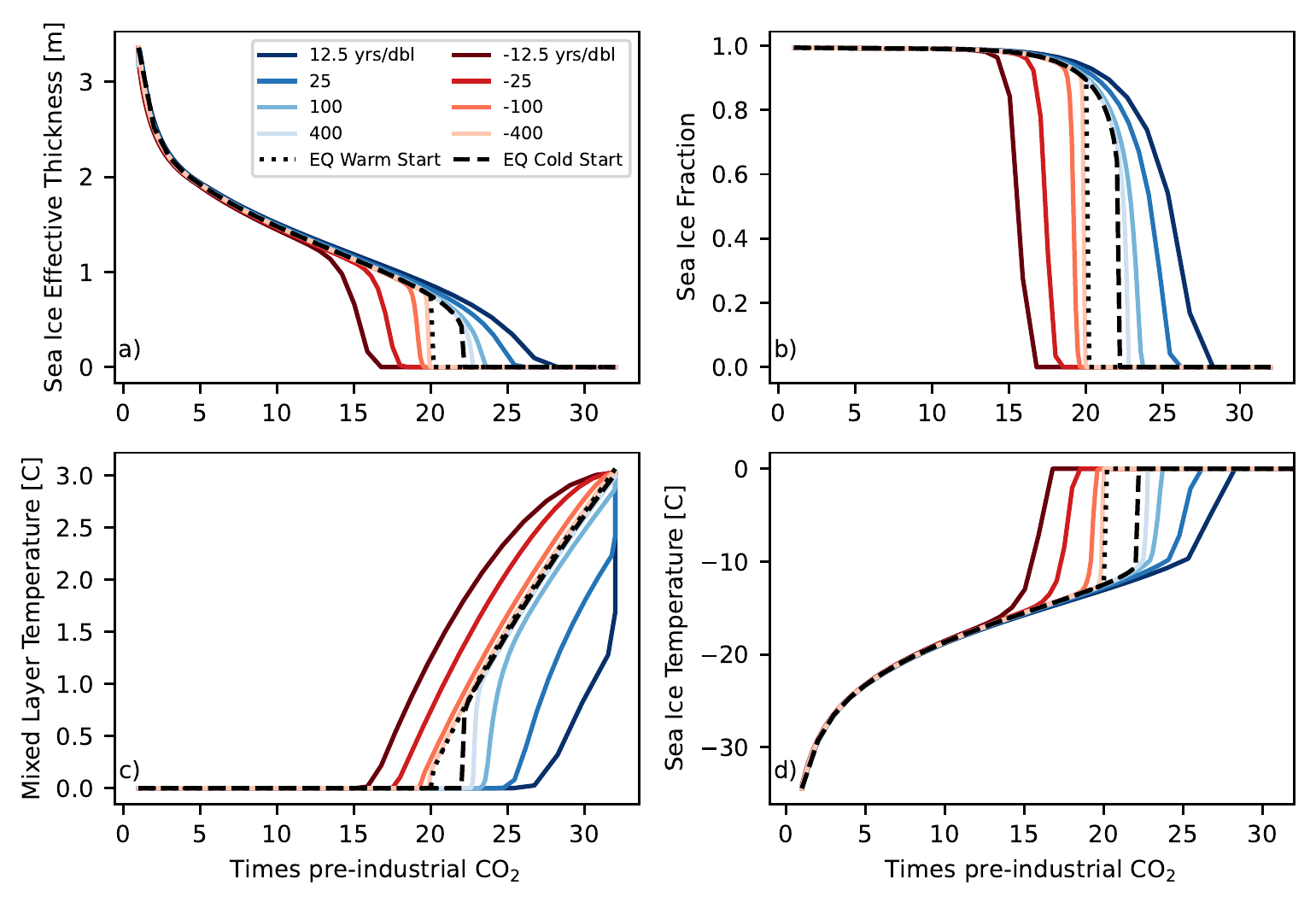}}
    \caption{\baselineskip=14pt Same as Figure S3, but for Scenario 2 (narrow bi-stability).}
\end{figure}

\begin{figure}
\centerline{
    \includegraphics[width=\textwidth]{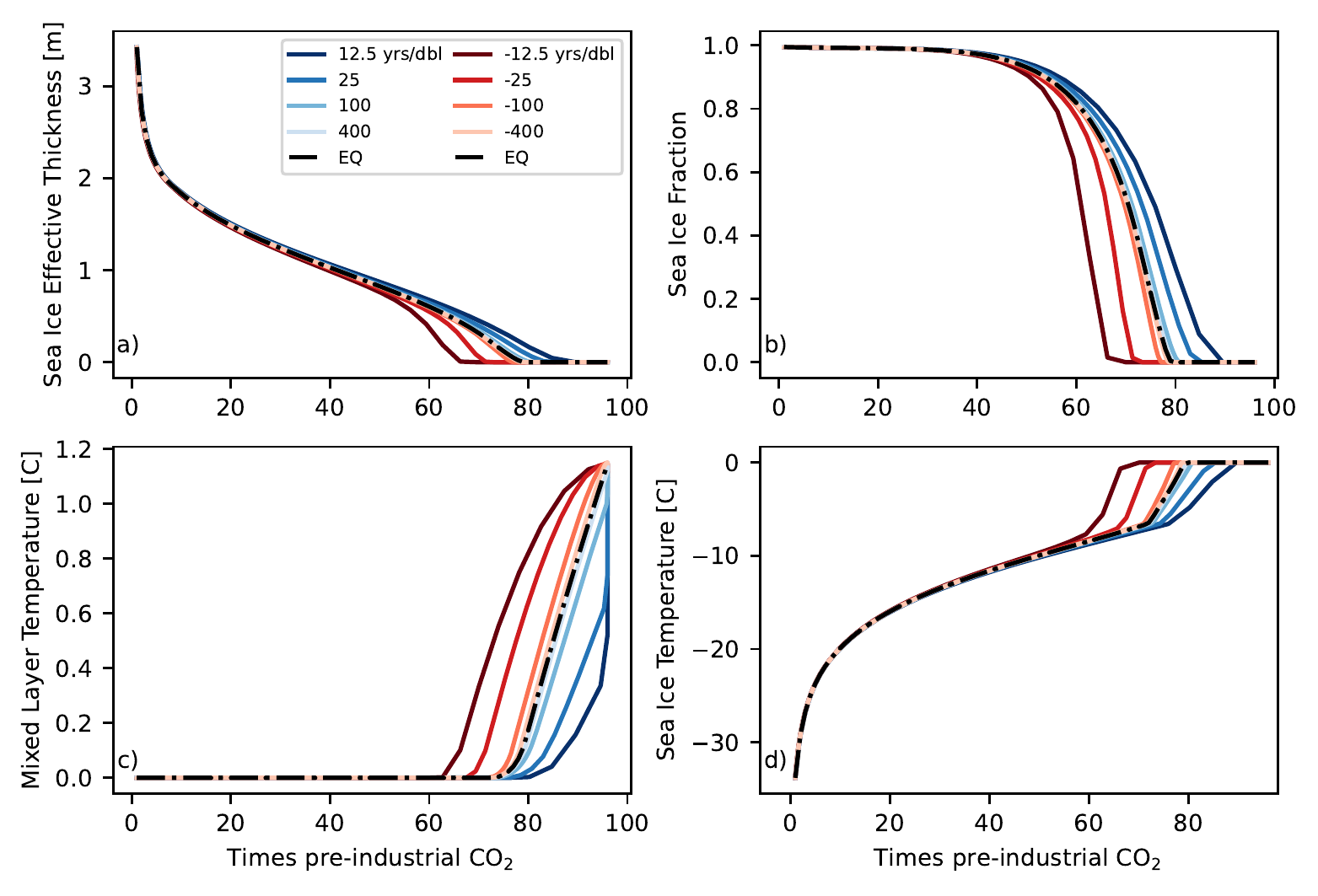}}
    \caption{\baselineskip=14pt Same as Figure S3, but for Scenario 3 (no bi-stability).}
\end{figure}

\begin{figure}
\centerline{
    \includegraphics[width=\textwidth]{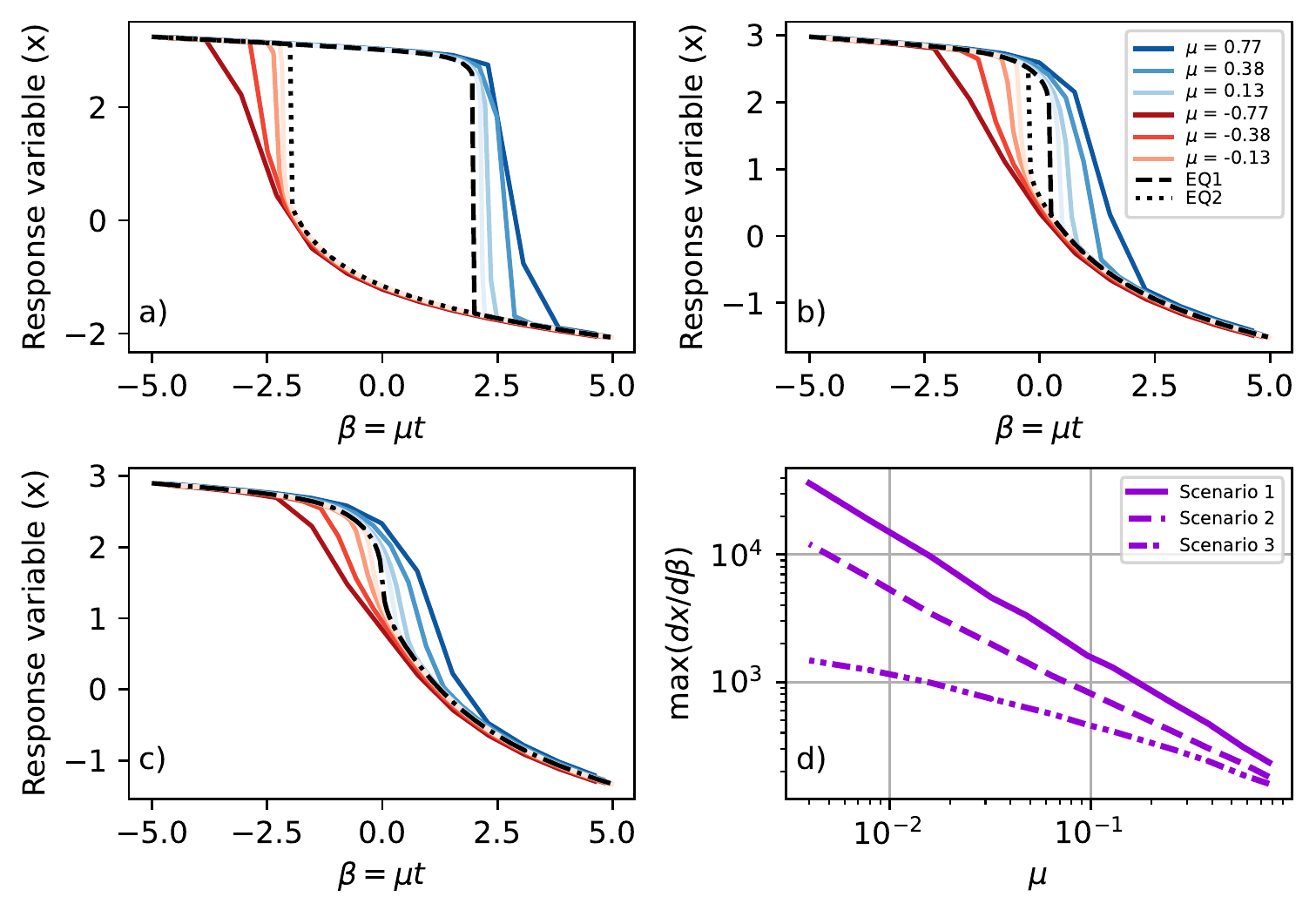}}
    \caption{\baselineskip=14pt Results from a version of the cubic ODE that is periodically forced (eq.~11). Panels a-c show transient (red and blue colored lines) and fixed-forcing (black lines) simulations for Scenarios 1, 2, and 3 respectively. Panel d shows the maximum gradient of $x$ with respect to the forcing parameter $\beta$ during transient simulations versus the ramping rate of each simulation for all three scenarios.  }
\end{figure}

\begin{figure}
\centerline{
    \includegraphics[width=\textwidth]{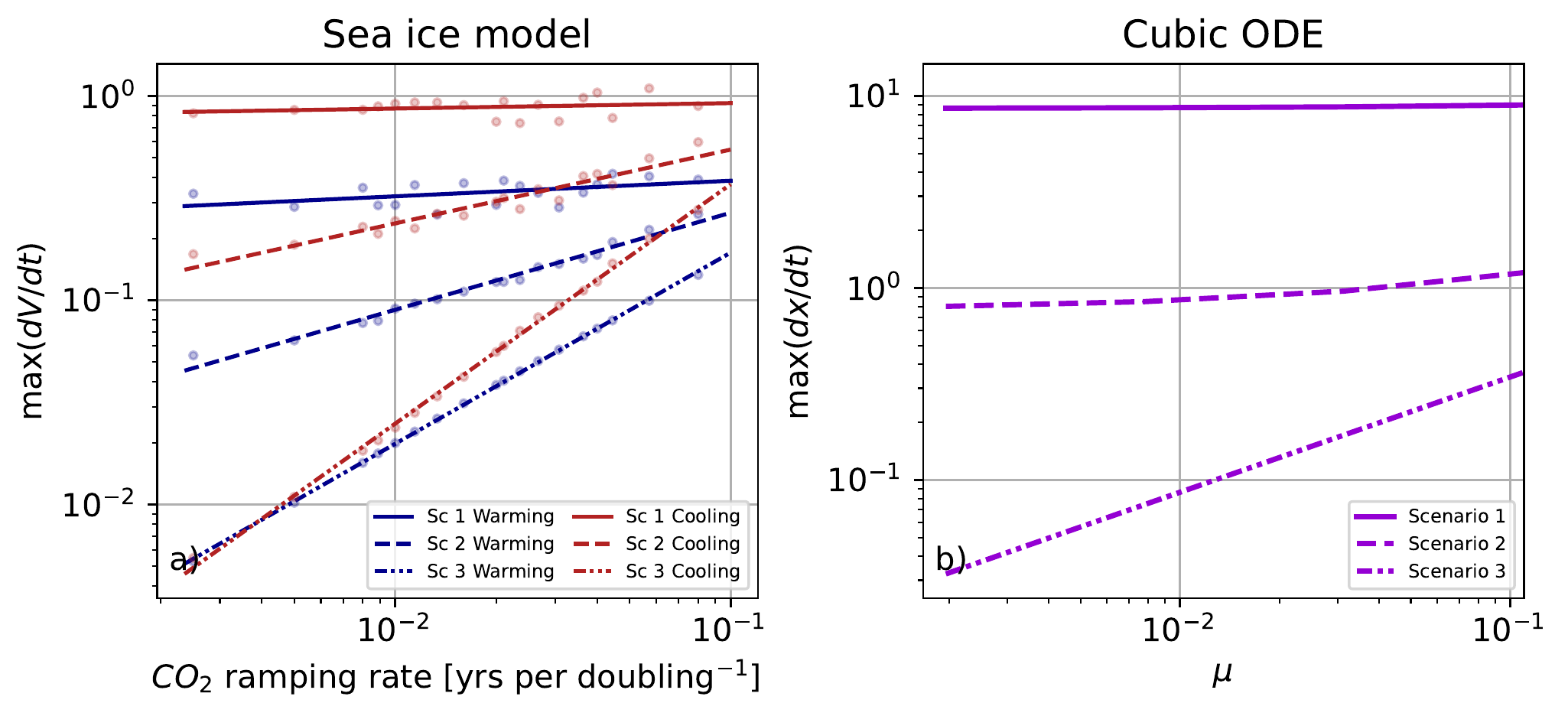}}
    \caption{\baselineskip=14pt Maximum rate of change of March sea ice effective thickness (a) and maximum rate of change of the variable $x$ from the cubic ODE (b) during time-dependent forcing simulations. }
\end{figure}

\begin{figure}
\centerline{
    \includegraphics[width=.8\textwidth]{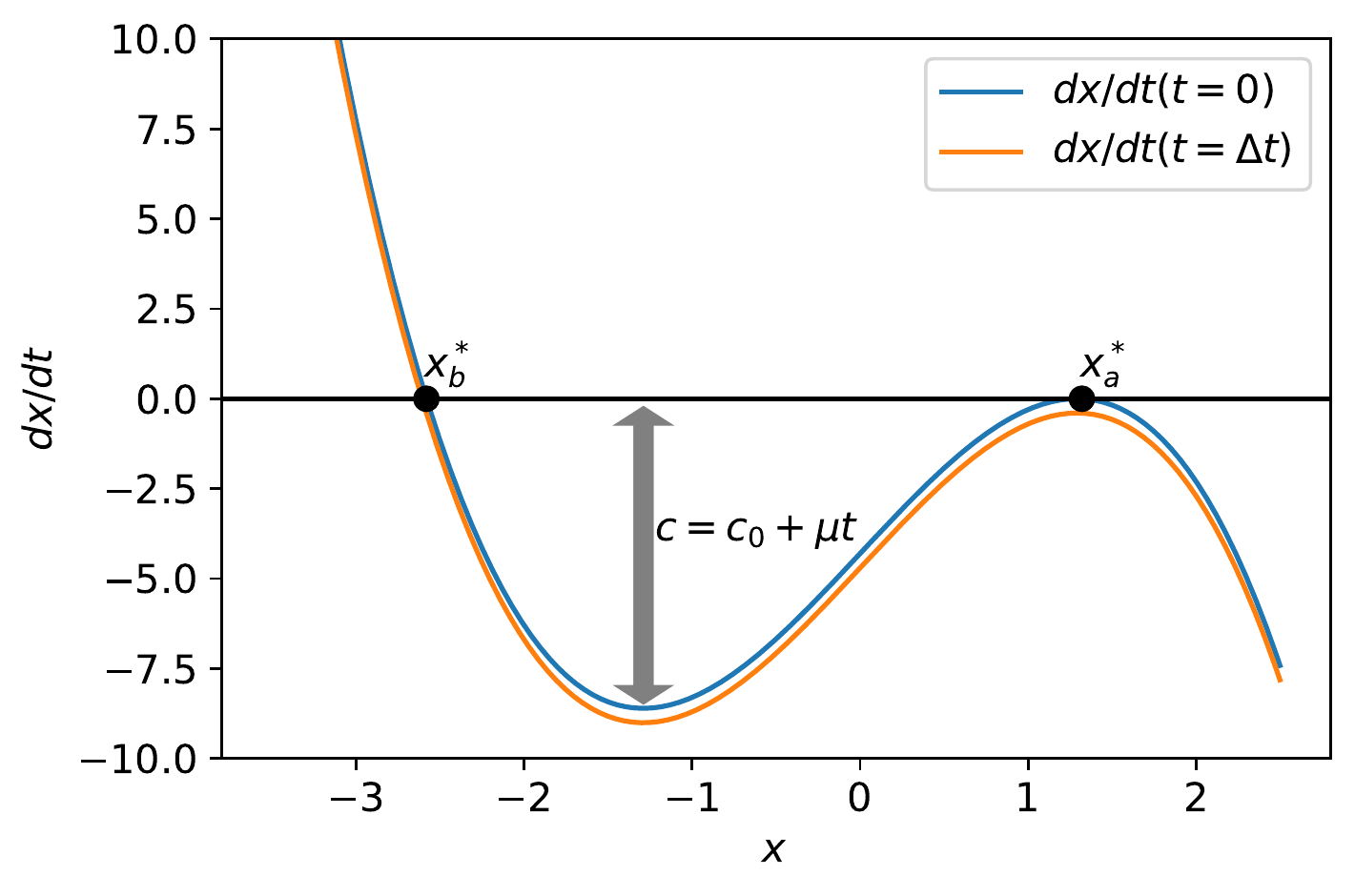}}
    \caption{\baselineskip=14pt Schematic of the upper limit on $\max(dx/dt)$ during the bifurcation for the equation $dx/dt=-x^3+5x-\mu t$ (Scenario 1 for the cubic ODE in the main text). The points $x_a^*$ and $x_b^*$ represent the two stable equilibria before the bifurcation; when the bifurcation happens, $x_a^*$ disappears and the solution must transition to $x_b^*$. The variable $c$, represents the upper limit on $\max(dx/dt)$, and $c_0$ is the value of $c$ at the time of the bifurcation, which is also the width of the true bi-stability.}
\end{figure}

\begin{figure}
\centerline{
    \includegraphics[width=.75\textwidth]{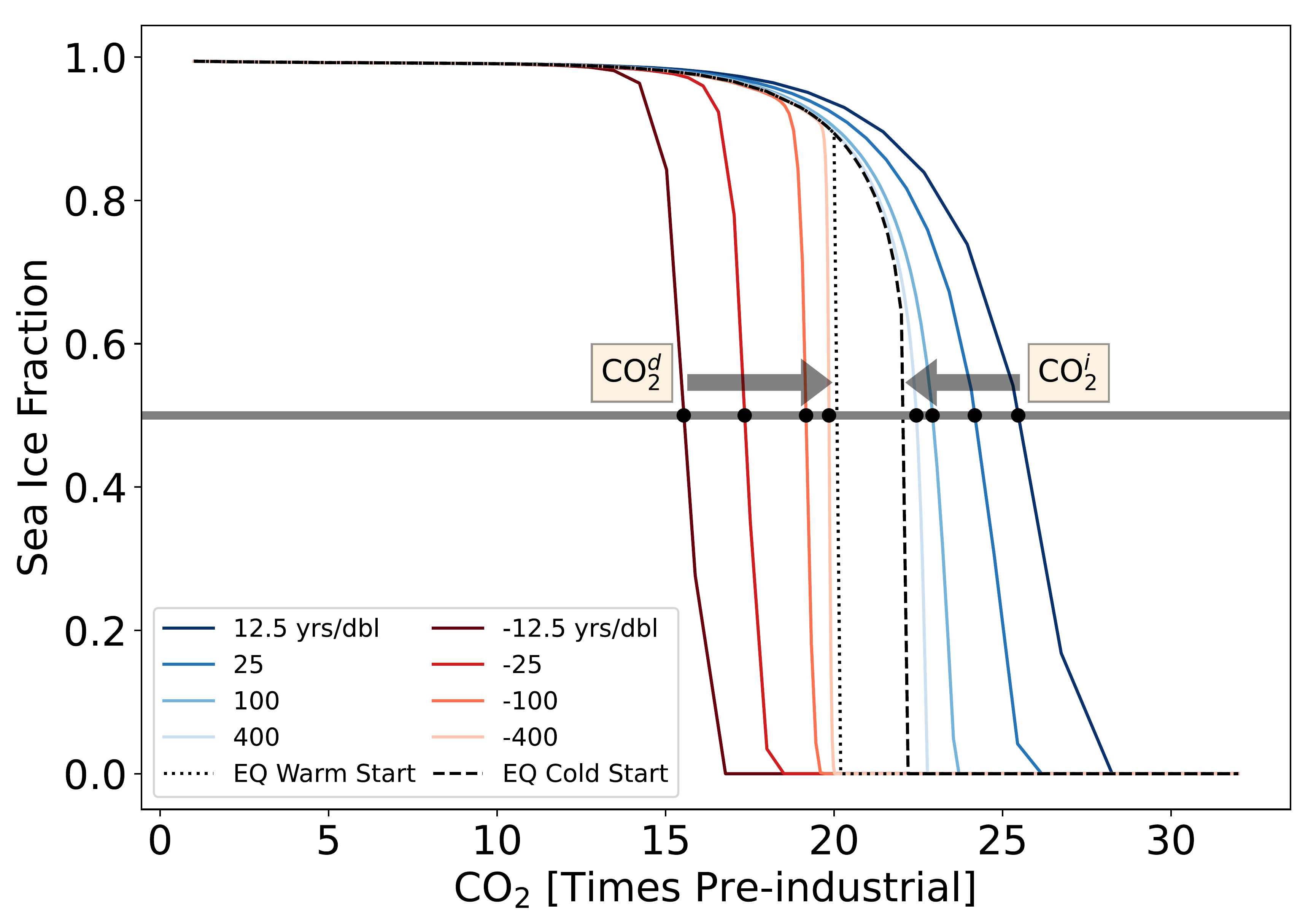}}
    \caption{\baselineskip=14pt Visualization of how the two edges of the transient hysteresis (CO$_2^i$ and CO$_2^d$) are calculated, demonstrated for Scenario 2 only. CO$_2^i$ and CO$_2^d$ are shown for a subset of the ramping rates as the block dots, and are the CO$_2$ values at which March ice fraction crosses a critical threshold (fraction of .5, shown in gray) along increasing and decreasing CO$_2$ forcing simulations respectively (see Methods in main text). As indicated by the gray arrows, we can see that for slower and slower ramping rates, CO$_2^i$ and CO$_2^d$ are converging to the width of bi-stability in the equilibrium.}
\end{figure}

%
% EXAMPLE LARGE TABLE (UPLOADED SEPARATELY)
%\begin{table}
%\settablenum{S1} %%Change number for each table
%\caption{Time of the Transition Between Phase 1 and Phase 2\tablenotemark{a}}
%\end{table}